\documentclass[showkeys,preprintnumbers,amsmath,amssymb,twocolumn,prl,floatfix]
{revtex4-1}
\usepackage{amssymb,amsfonts}
\usepackage{amsmath}
\usepackage{hyperref}
\usepackage{graphicx}
\usepackage{natbib}
\usepackage[usenames,dvipsnames]{xcolor}
\graphicspath{{Figures/}}
\newcommand{\one}{\mbox{$1 \hspace{-1.0mm}  {\bf l}$}}

\begin{document}
\title{Decoherence as Detector of the Unruh Effect}



\author{Alexander I Nesterov}
  \email{nesterov@cencar.udg.mx}
\affiliation{Departamento de F{\'\i}sica, CUCEI, Universidad de Guadalajara,
Guadalajara, CP 44420, Jalisco, M\'exico}

\author{Gennady P  Berman}
 \email{gpb@lanl.gov}
\affiliation{Theoretical Division, Los Alamos National Laboratory,  Los Alamos, NM 87545, USA}

\author{Manuel A Rodr\'iguez Fern\'andez}
  \email{mda.rzfz@gmail.com}
\affiliation{Departamento de F{\'\i}sica, CUCEI, Universidad de Guadalajara,
Guadalajara, CP 44420, Jalisco, M\'exico}

\author{Xidi Wang}
 \email{xidiwang@ucsd.edu}
 \affiliation{Department of Chemistry ${\&}$ Biochemistry, University of California San Diego, La 
 Jolla CA 92093, USA}
 
\date{\today}

\begin{abstract}
We propose a new type of the Unruh-DeWitt detector which  measures the decoherence of the 
reduced density matrix of the detector interacting with the massless quantum  scalar field. We 
find that the decoherence decay rates are different in the inertial and accelerated reference 
frames. We show that the exponential phase decay  can be observed for 
relatively low accelerations, that can significantly improve the conditions for measuring the 
Unruh effect.
\end{abstract}


\keywords{Effect Unruh, Rindler spacetime, scalar field, decoherence}

\preprint{LA-UR-20-22227}

\maketitle
All elementary particles, which exist in nature, are the excitations of the corresponding quantum 
fields. But even when these fields are in their quantum vacuum states, a very complicated 
dynamics of vacuum fluctuations takes place.  Moreover, as it was discovered by Unruh in 1976  
\cite{UWG}, in a flat spacetime the state of the 
 quantum vacuum  depends on the motion of observer, and thus the concept of particles, in the 
 context of quantum field theory, is relative. More specifically, for the uniformly accelerated 
 observer the vacuum of quantum fields in the Minkowski spacetime is modified to the thermal 
 state. This result,  nowadays known  as the Unruh effect, generated an enormous number of 
 publications. (See Ref. \cite{CLH} for a comprehensive review,  and references therein.) The 
 effective temperature of the thermal distribution is $T =\hbar a/2\pi c k_B$, where  $a$  is the 
 acceleration, $k_B$  is the Boltzmann constant,  $\hbar$  is the reduced Planck constant, and  
 $c$  is the speed of light. 

 The direct observation of the Unruh effect requires very large accelerations, e. g. $a \approx 2.5 \cdot 10^{20} \, \rm m/s^2$  for $T = 1 K$. These accelerations can be realized only in strong gravitational fields, for instance, produced by the black holes. That is why, the current trend in this field is to propose various types of detectors which can measure indirectly the Unruh effect \cite{U95,UW,HDMM,ZWPR,FHY}. In particular, the detector which measures the dependence of the Berry phase on the acceleration was suggested in \cite{MMFE,MMED}. According to Ref. \cite{MMFE}, this detector can significantly reduce the required acceleration, up to $a \approx 2.5 \cdot 10^{17} \, \rm m/s^2$.

In this Letter, we propose a new type of detector to probe the  Unruh effect. Our detector 
does not measure directly the equilibrium distribution of the produced particles, but the 
decoherence  of the reduced density matrix.  We show that this information can be used for the 
detection of the Unruh effect at significantly lower accelerations, even in comparison with the 
detector based on the measurement of the Berry phase. Through the paper, unless stated 
otherwise, we use the natural units, $\hbar =c =k_B =1$. \\

We assume that the observer moves with an uniform acceleration, $a$, in $z$-direction with respect to an inertial reference frame in the Minkowski spacetime. 
The transformation of coordinates,
\begin{align}
t = \frac{1}{a} e^{a\zeta} \sinh(a\tau), \quad
z = \frac{1}{a} e^{a\zeta} \cosh(a\tau),
\end{align}
describes the right wedge of the Rindler spacetime with the metric,
\begin{align}
	ds^2 = e^{2a\zeta}(d\tau^2 - d\zeta^2) - d \textbf{x}_{\bot}^2,
\end{align}
 where we set ${\mathbf x}_\perp = (x,y)$. Hereafter, we denote the coordinates in the Minkowski spacetime as $(t, \mathbf r)$, and the coordinates in the Rindler spacetime as $(\tau,\mathbf x )$, where  $ \mathbf x  =({\mathbf x}_\perp, \zeta )$.

 The conventional Unruh-DeWitt (UDW) detector is presented as a box containing a non-relativistic particle interacting with a massless scalar field. It is assumed that the detector is located at the origin of the moving reference frame, and the particle is in its ground state. The quanta of the scalar field is detected if the detector is found in an excited state. Since, only two states are relevant: the ground state and the first excited state, one can consider the detector as a two-level quantum  system interacting with the scalar field \cite{UWG,UW}. 

The entire system, ``detector + field", is governed by the Hamiltonian, 
$\mathcal H_t = H_d +H_f + H_{int}$, where $H_d $ is the Hamiltonian of the detector, $H_f$ is the  Hamiltonian of the free scalar field{\tiny }, and $H_{int}$ stands for the interaction Hamiltonian. The latter can be written as,
$H_{int}= \int_{\Sigma_t} {\mathcal H}_{int} \,d^3 \mathbf r$,
where ${\mathcal H}_{int}$ is the Hamiltonian density, and the integral is taken over the three-dimensional surface, $\Sigma_t$, at time $t = \rm const$. 

In the original formulation, of the UDW model, the detector was considered as a pointlike 
particle, with the interaction Hamiltonian being $H_{int}= \int_{\Sigma_t} \delta^3(\mathbf r- 
\mathbf r(t)){\mathcal H}_{int} \,d^3 \mathbf r$, 
where $\mathbf r (t)$, describes the trajectory of the detector. In our paper we consider both 
cases:  a pointlike detector and a detector of the finite size. 

In our analysis, we consider the UDW detector
as a two-level system with the transition  energy $\varepsilon$. The Hamiltonian of the 
detector we take as $H_d= (\varepsilon /2)\sigma_z $, where $\sigma_z$ is the 
Pauli matrix. This form of the Hamiltonian corresponds to the effective Zeeman interaction of a 
single spin with a permanent magnetic field, oriented in the $z$-direction. 

 We assume that in the uniformly accelerated reference frame, the  detector is located at the 
 origin of the coordinates. The entire system is governed by the Hamiltonian,
\begin{align}
&H_{t} = {{\varepsilon}\over{2}}\sigma_z \otimes \one + \one \otimes\sum_{\omega > 0 
}\sum_{ \textbf{k}_{\bot}} 
\omega b^\dagger_{\mathbf k} b_{\mathbf k}  \nonumber \\
&+\frac{\lambda}{\sqrt{V}}\sigma_z \otimes\sum_{\omega > 0 
}\sum_{ \textbf{k}_{\bot}}  \frac{1 }{\sqrt{2\omega}}(g_{\mathbf k} b^\dagger_{\mathbf k}+ 
g^\ast_{\mathbf k} b_{\mathbf k}),
\label{H}
\end{align}
 where $\sigma_z$ is the Pauli matrix, $\lambda$ is a coupling constant and $\mathbf{k} = 
  (\textbf{k}_{\bot}, \omega)$. The Rindler 
 creation and  annihilation operators, associated with the field mode, $\mathbf{k}$, obey the 
 standard commutation relations: $[b_{\mathbf k}, 
 b^\dagger_{{\mathbf k}'}] = \delta_{{\mathbf k}{\mathbf k}'}$, $[b_{\mathbf k}, b_{{\mathbf 
  k}'}] = 0$ and $[b^\dagger_{\mathbf k}, b^\dagger_{{\mathbf k}'}] = 0$. The form factor, $ 
  g_{\mathbf k}$, 
 is defined as follows: $ g_{\mathbf k}= \int_{\mathbf 
	R^3}v_{\mathbf k} ( {\mathbf x})f(\mathbf x) e^{a\zeta} d \zeta\,d^2 \mathbf x_\perp$. The modes, $v_{\mathbf 	k} $,  are given by \cite{SAFNCQ,FSA1,LFOK},
\begin{align}
v_{\mathbf k} = \sqrt{\frac{4 \omega\sinh(\pi \omega/a)}{\pi a}}K_{i \omega / a} \Big( 
\frac{k_{\bot}}{a} e^{a\zeta} \Big )e^{i \mathbf{k}_{\perp}\cdot \mathbf{x}_{\perp}} ,
\end{align}
where $K_{i \mu}(z)$ is the Macdonald 
 function of the {\em imaginary} order. The function, $f(\mathbf x) $, describes the spatial 
 profile of the detector in the Rindler space.
 
In \eqref{H}, we assume that the interaction term is much smaller than the effective Zeeman 
interaction. It is well-known that in this case, only the $\sigma_z$ operator can be used in the 
interaction Hamiltonian. We call the system with the Hamiltonian (\ref{H}) energy conserving, 
because the operator, 
	$\sigma_z$, commutes with the total Hamiltonian.  As a result, the initial probabilities 
	of population of the detector do not change in time.

 We assume  that for the entire system, the density operator, $\varrho(\tau)$,  at time  	
$\tau=\tau_0$, takes the form, $\varrho(\tau_0) = |\Psi_0\rangle\langle\Psi_0 |$, with 
 	$|\Psi_0\rangle = |\psi_0\rangle\otimes |0_M \rangle$.  Here, 
 	$|\psi_0\rangle=\alpha|\downarrow\rangle+\beta |\uparrow\rangle$, denotes the  	initial 
 	{\em{superpositional}} state of the detector, and $|0_M \rangle$ stands for the Minkowski 
 	vacuum.

We denote by $\rho(\tau)$ the detector reduced  density matrix, obtained by tracing out all 
scalar field degrees of freedom. The {time evolution of the} matrix elements  of the reduced 
density matrix can be written as,
\begin{align} \label{EqR}
\rho_{ij}(\tau) = & \langle i| {\rm Tr}_R U(\tau)\varrho(0) U^{-1}(\tau)|j \rangle, \\
& (i,j=0,1), \nonumber
\end{align}
where the index $i=0$ is associated with the eigenvector $|\downarrow \rangle$, and the 
index $i=1$ is associated with the eigenvector $|\uparrow \rangle$ of  the operator 
$\sigma_z$.

The interaction of the detector with the scalar field does not excite the detector, and the 
detection of the Unruh effect is reduced to the study of the phase decoherence (decay of the 
non-diagonal elements of the reduced density matrix) (for details see the supplemental material 
(SM) at https:// \dots ):
\begin{align}
\rho_{01} (\tau)= e^{i \varepsilon \tau-\gamma(\tau)}\rho_{01} (\tau_0).
\end{align} 
 We say that the {\em full} phase decoherence takes place if $\rho_{01}(\tau) \rightarrow 0$ as 
 $\tau \rightarrow \infty$. Otherwise, we call the phase 
decoherence {\em partial}.

 The computation of the decoherence function yields  (see SM):
\begin{align}
\gamma (\tau)=\frac{\lambda^2}{2\pi^3} \int_0^\infty \frac{|g_\omega|^2}{\omega}{\rm 
sinc}^2 \Big(\frac{\omega \tau}{2} \Big )\coth\Big(\frac{\pi \omega}{a}\Big) d\omega,
\label{D5g}
\end{align}
where $|g_\omega |^2= \int |g_{\mathbf k}|^2 d^2 {\mathbf k}_\perp $, ${\rm sinc} (x) =\sin (x)/x$. \\

{\em Pointlike detector.} --  For a pointlike detector, the form factor is, 
\begin{align}
g_{\mathbf k} = \sqrt{\frac{4 \omega\sinh(\pi \omega/a)}{\pi a}}K_{i \omega / a} \Big( 
\frac{k_{\bot}}{a} \Big ).
\end{align}
The  computation of  $ |g_\omega|^2$ yields, $ |g_\omega|^2 = 8 \pi\omega^{2}$. 
Substituting this result in Eq. \eqref{D5g}, we obtain,
\begin{align}
\gamma (\tau)=\frac{4\lambda^2}{\pi^2} \int_0^\infty {\omega} \,{\rm sinc}^2 
\Big(\frac{\omega \tau}{2} \Big )\coth\Big(\frac{\pi \omega}{a}\Big) d\omega.
\label{D5b}
\end{align}

As one can see, the integral in (\ref{D5b}) is formally divergent at $\omega \rightarrow 
\infty$, and  we have, $\gamma \propto \ln \omega$. This issue was studied in  
\cite{SBS,SBS1,HAM,SLPT}. It was shown that the ultraviolet logarithmic divergence is 
caused by instantaneous switching on/off  of the detector.  This difficulty can be overcome, and 
the divergence can be removed by a regularization procedure through the smooth switching 
function of the detector, or through its profile (or 
both) \cite{HDMM,LCBF}. Below we use the regularization procedure through the detector 
profile.\\

{\em Detector of the finite size.} --  To avoid the ultraviolet divergence, we consider the form factor in the form,
\begin{align}
g_{\mathbf k} = e^{-l\omega/2}\sqrt{\frac{4 \omega\sinh(\pi \omega/a)}{\pi a}}K_{i \omega / 
a} \Big( \frac{k_{\bot}}{a} \Big ) ,
\label{Ff}
\end{align}
where, $l$, is the characteristic size of the detector. The exponential cutoff eliminates the 
logarithmic divergence in the limit of $\omega \rightarrow \infty$. Using the inverse 
transformation,  one can reconstruct the detector profile as follows:
\begin{align}
f(\mathbf x) = \frac{1}{(2\pi)^3}  \int g_{\mathbf k}v^\ast_{\mathbf k} ( {\mathbf x})\,d^3 \mathbf k.	
\end{align}

For the detector at rest in Minkowski space, the computation yields the spherically 
symmetric profile:
\begin{align}
	f(r) = \frac{l}{\pi^2(l^2 + r^2)^2}.
\end{align}
 One can show, that at $l \rightarrow 0$, the function $f(r) \rightarrow \delta (\mathbf r)$. Thus, the detector becomes a pointlike particle, and we return to the expression \eqref{D5b} for the decoherence function. 
 
 The choice of the magnitude of the cutoff is dictated by the minimal size of the detector. We assume here that it can't be less than the size of the hydrogen atom, and set $l \approx r_0$, where $r_0$ is the Bohr radius.

Due to the acceleration, in the Rindler spacetime the detector modifies its shape  as follows:
\begin{align}
	f(\mathbf x) = \frac{l e^{-a \zeta} \ln(u +\sqrt{u^2 -1} )}{\pi^2 \sqrt{u^2 -1} \big(l^2 + \frac{1}{a^2} \ln^2(u +\sqrt{u^2 -1} )\big)^2 },
\end{align}
where $u= a^2(x^2 + y^2) e^{-a \zeta} /2 + \cosh(a \zeta)$. 

\begin{figure}[tbh]
\scalebox{0.275}{\includegraphics{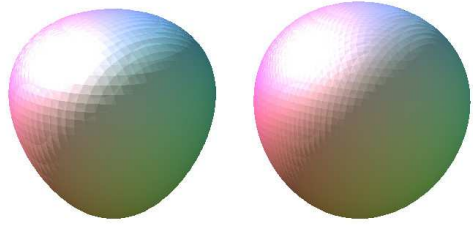}}
\caption{(Color online) The detector profile. Left: an accelerated observer, with the acceleration being taken as, $a = 5 \cdot 10^{26} \rm m/s^2$. Right: an inertial observer ($a=0$).
\label{fig1}}
\end{figure}

In Fig. \ref{fig1}, we compare the detector shape in the inertial reference frame with its shape in the accelerated reference frame, for the  observer moving with the acceleration, $a = 5 \cdot 10^{26} \rm m/s^2$. Such high acceleration can be sustained by using current laser technologies \cite{CLH,CPT}.

With the modified form factor, the decoherence function takes the form,
\begin{align}
\gamma (\tau)=\frac{4\lambda^2}{\pi^2} \int_0^\infty {\omega} e^{-l\omega } {\rm sinc}^2 
\Big(\frac{\omega \tau}{2} \Big )\coth\Big(\frac{\pi \omega}{a}\Big) d\omega.
\label{D6a}
\end{align}

Performing the integration, we obtain,
\begin{align}
\gamma(\tau) = &\frac{2\lambda^2 }{\pi^2}\Re\ln (1 +{i \tau}/l)   \nonumber \\
&-\frac{4\lambda^2 }{\pi^2}\Re \ln\bigg (\frac{\Gamma (1 +al/2\pi +i  a\tau /2\pi )}{\Gamma (1 + al/2\pi)}\bigg),
\label{Zeta1}
\end{align}
where $\Gamma(z)$ is the Gamma-function. 
\begin{figure}[tbh]
\scalebox{0.36}{\includegraphics{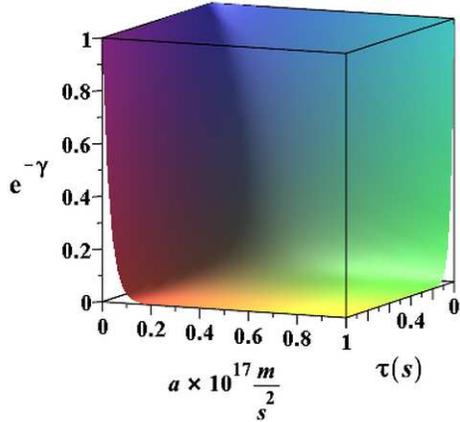}}
\caption{(Color online) The exponential decay function, $e^{-\gamma(\tau)}$ vs $\tau$ and $a$ ($\lambda = 10^{-3}$).
\label{Fig3a}}
\end{figure}
\begin{figure}[tbh]
\begin{center}
\scalebox{0.28}{\includegraphics{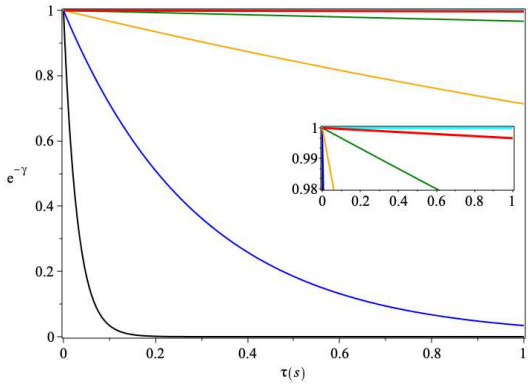}}
\end{center}
\caption{(Color online) The exponential decay function, $e^{-\gamma(\tau)}$ vs $\tau$ ($\lambda =10^{-3}$). From the top to the bottom: $a=0$  (cyan),  $a=10^{13}\rm  m/s^2$ (red), $a=10^{14}\rm  m/s^2$ (green),  $a=10^{15}\rm  m/s^2$ (orange), $a=10^{16}\rm  m/s^2$ (blue), $a=10^{17}\rm  m/s^2$  (black). Inset: zoom of the main figure.
\label{Fig2b}}
\end{figure}

Returning to the physical units, we find that in the limit of $ a \tau/2\pi c\gg 1$, one can approximate the decoherence function as, 
\begin{align}
\gamma(\tau) \approx \frac{2\lambda^2 }{\pi^2}\ln (c\tau/l)  + \frac{\lambda^2 a\tau}{\pi^2 c} .
\label{Gamma}
\end{align}
 As one can see, the detector exhibits the full  phase decoherence, even for an inertial motion.  
 However, the decay process is very slow, $\gamma \propto \ln \tau $.
  In the limit of  large accelerations, one can neglect the contribution of the first term in Eq. \eqref{Gamma}, and recast the decoherence function as, $\gamma(\tau) \approx \Gamma_d \tau $, where $\Gamma_d =\lambda^2 a/(\pi^2 c) $ denotes the decoherence decay rate. We conclude that the decoherence effect is insensitive to the choice of the cutoff and is  highly sensitive to the choice of the coupling constant,  $\lambda$.
 
 In Figs. \ref{Fig3a}, and \ref{Fig2b}, we present the results of numerical simulations for the trial coupling constant, $\lambda = 10^{-3}$.  In Fig. \ref{Fig3a}, the exponential decay function is depicted as a function of the acceleration and proper time of the accelerating observer. In Fig. \ref{Fig2b}, we plot the exponential decay function as a function of time and for different choices of the acceleration: $a=0$, $a=10^{13} \rm  m/s^2$, $a=10^{14} \rm m/s^2$,  $a=10^{15} \rm m/s^2$, $a=10^{16} \rm m/s^2$ and $a=10^{17}\rm  m/s^2$.\\

{\em Setup for the Gedanken experiment.} -- We consider two identical detectors coupled to 
the scalar field in both inertial and accelerated reference frames. The experiment consists of 
comparing the decoherence for inertial and accelerated observers.

Let us denote by $\sigma$ the magnitude of the decoherence function that can be measured in 
the experiment. Then, the time  required to make the measurement, can be estimated as 
follows: $\tau_i \approx (l/c )\exp(\pi^2 \sigma/\lambda^2) $  for the 
detector at rest in Minkowski space, and $\tau_a =\sigma /\Gamma_d $ for the uniformly 
accelerated detector. By choosing, $\sigma = 
10^{-4}$, we find $\tau_i \sim e^{10^3} \,\rm s$, and $\tau_a \approx 3 \cdot 10^{11}/a 
\,[\rm s] $. Then, for the acceleration, $a= 10^{13} \rm m/s^2$, we obtain, $\tau_a \approx 
30 \,\rm ms$, and $\tau_a \approx 3 \,\rm \mu s$ for $a= 10^{17} \rm m/s^2$.

{\em Laboratory bounds.} -- Due to the Lorentz contraction of time, for the high accelerations, one might expect a significant difference between the decoherence time for the accelerated observer and for the inertial one. Indeed, while for non-relativistic velocities, the observation time in the laboratory system is, $\Delta t \approx \tau$, where $\tau$ denotes the decoherence time for the accelerated observer, for the  ultra-relativistic motion, one might expect, $\Delta t \gg \tau$. This imposes strong restrictions on the upper limit of the observation time in the Rindler spacetime.

For definiteness, let us consider two cases: (1) non-relativistic motion with $v\lesssim 0.2 \,c $; (2) ultra-relativistic motion with $v\lesssim 0.95 \,c $. In the first case, we obtain: $\Delta t_1 \approx \tau_1$, and $\tau_1 \lesssim 0.2 c/a$. In the second case, we find: $\Delta t _2\approx 1.8\tau_2$ and $\tau_2 \lesssim  2c/a$. Substituting $\tau_{1,2} $ in Eq. \eqref{Gamma}, we obtain $\gamma(\tau_1) \approx 0.02 {\lambda^2 }$, and $\gamma(\tau_2) \approx 0.2 {\lambda^2 }$. \\

{\em Concluding remarks.} --  We demonstrated that the information about the presence of the 
Unruh effects is encoded in the 
decoherence - the  exponential decay of the non-diagonal elements of the reduced density 
matrix. The phase decay of the reduced density matrix can be observable for the 
accelerations as low as $10^{13}\, \rm m/s^2$. Such accelerations should be sustained only for 
a time $\approx 100 \, \mu s$.  

Our approach has the limitations related to the decoherence observation time from the perspective of an inertial observer at rest, in the Minkowski spacetime.  Note, that the same restrictions are valid for any detector based on the indirect measurement of the Unruh effect.
\\

\begin{acknowledgements}

The work by G.P.B. was done at Los Alamos National Laboratory  managed by Triad National Security, LLC, for the National Nuclear Security Administration of the U.S. Department of Energy under Contract No. 89233218CNA000001. A.I.N. and  M.A.R.F. acknowledge the support by CONACyT (Network Project No. 294625, ``Agujeros Negros y Ondas Gravitatorias'').

\end{acknowledgements}


\begin{thebibliography}{20}%
\makeatletter
\providecommand \@ifxundefined [1]{%
 \@ifx{#1\undefined}
}%
\providecommand \@ifnum [1]{%
 \ifnum #1\expandafter \@firstoftwo
 \else \expandafter \@secondoftwo
 \fi
}%
\providecommand \@ifx [1]{%
 \ifx #1\expandafter \@firstoftwo
 \else \expandafter \@secondoftwo
 \fi
}%
\providecommand \natexlab [1]{#1}%
\providecommand \enquote  [1]{``#1''}%
\providecommand \bibnamefont  [1]{#1}%
\providecommand \bibfnamefont [1]{#1}%
\providecommand \citenamefont [1]{#1}%
\providecommand \href@noop [0]{\@secondoftwo}%
\providecommand \href [0]{\begingroup \@sanitize@url \@href}%
\providecommand \@href[1]{\@@startlink{#1}\@@href}%
\providecommand \@@href[1]{\endgroup#1\@@endlink}%
\providecommand \@sanitize@url [0]{\catcode `\\12\catcode `\$12\catcode
  `\&12\catcode `\#12\catcode `\^12\catcode `\_12\catcode `\%12\relax}%
\providecommand \@@startlink[1]{}%
\providecommand \@@endlink[0]{}%
\providecommand \url  [0]{\begingroup\@sanitize@url \@url }%
\providecommand \@url [1]{\endgroup\@href {#1}{\urlprefix }}%
\providecommand \urlprefix  [0]{URL }%
\providecommand \Eprint [0]{\href }%
\providecommand \doibase [0]{http://dx.doi.org/}%
\providecommand \selectlanguage [0]{\@gobble}%
\providecommand \bibinfo  [0]{\@secondoftwo}%
\providecommand \bibfield  [0]{\@secondoftwo}%
\providecommand \translation [1]{[#1]}%
\providecommand \BibitemOpen [0]{}%
\providecommand \bibitemStop [0]{}%
\providecommand \bibitemNoStop [0]{.\EOS\space}%
\providecommand \EOS [0]{\spacefactor3000\relax}%
\providecommand \BibitemShut  [1]{\csname bibitem#1\endcsname}%
\let\auto@bib@innerbib\@empty
\bibitem [{\citenamefont {Unruh}(1976)}]{UWG}%
  \BibitemOpen
  \bibfield  {author} {\bibinfo {author} {\bibfnamefont {W.~G.}\ \bibnamefont
  {Unruh}},\ }\bibfield  {title} {\enquote {\bibinfo {title} {{Notes on
  black-hole evaporation}},}\ }\href {\doibase 10.1103/PhysRevD.14.870}
  {\bibfield  {journal} {\bibinfo  {journal} {Phys. Rev. D}\ }\textbf {\bibinfo
  {volume} {14}},\ \bibinfo {pages} {870--892} (\bibinfo {year}
  {1976})}\BibitemShut {NoStop}%
\bibitem [{\citenamefont {{L. C. B. Crispino, A. Higuchi and G. E. A.
  Matsas}}(2008)}]{CLH}%
  \BibitemOpen
  \bibfield  {author} {\bibinfo {author} {\bibnamefont {{L. C. B. Crispino, A.
  Higuchi and G. E. A. Matsas}}},\ }\bibfield  {title} {\enquote {\bibinfo
  {title} {{The Unruh effect and its applications}},}\ }\href {\doibase
  10.1103/RevModPhys.80.787} {\bibfield  {journal} {\bibinfo  {journal} {Rev.
  Mod. Phys.}\ }\textbf {\bibinfo {volume} {80}},\ \bibinfo {pages} {787--838}
  (\bibinfo {year} {2008})}\BibitemShut {NoStop}%
\bibitem [{\citenamefont {{W. G. Unruh}}(1995)}]{U95}%
  \BibitemOpen
  \bibfield  {author} {\bibinfo {author} {\bibnamefont {{W. G. Unruh}}},\
  }\bibfield  {title} {\enquote {\bibinfo {title} {{Maintaining coherence in
  quantum computers}},}\ }\href {\doibase 10.1103/PhysRevA.51.992} {\bibfield
  {journal} {\bibinfo  {journal} {Phys. Rev. A}\ }\textbf {\bibinfo {volume}
  {51}},\ \bibinfo {pages} {992--997} (\bibinfo {year} {1995})}\BibitemShut
  {NoStop}%
\bibitem [{\citenamefont {{W. G. Unruh and R. M. Wald}}(1984)}]{UW}%
  \BibitemOpen
  \bibfield  {author} {\bibinfo {author} {\bibnamefont {{W. G. Unruh and R. M.
  Wald}}},\ }\bibfield  {title} {\enquote {\bibinfo {title} {{What happens when
  an accelerating observer detects a Rindler particle}},}\ }\href {\doibase
  10.1103/PhysRevD.29.1047} {\bibfield  {journal} {\bibinfo  {journal} {Phys.
  Rev. D}\ }\textbf {\bibinfo {volume} {29}},\ \bibinfo {pages} {1047--1056}
  (\bibinfo {year} {1984})}\BibitemShut {NoStop}%
\bibitem [{\citenamefont {{D. H\"ummer, E. Mart\'{\i}n-Mart\'{\i}nez and A.
  Kempf}}(2016)}]{HDMM}%
  \BibitemOpen
  \bibfield  {author} {\bibinfo {author} {\bibnamefont {{D. H\"ummer, E.
  Mart\'{\i}n-Mart\'{\i}nez and A. Kempf}}},\ }\bibfield  {title} {\enquote
  {\bibinfo {title} {{Renormalized Unruh-DeWitt particle detector models for
  boson and fermion fields}},}\ }\href {\doibase 10.1103/PhysRevD.93.024019}
  {\bibfield  {journal} {\bibinfo  {journal} {Phys. Rev. D}\ }\textbf {\bibinfo
  {volume} {93}},\ \bibinfo {pages} {024019} (\bibinfo {year}
  {2016})}\BibitemShut {NoStop}%
\bibitem [{\citenamefont {{W. Zhou, R. Passante and L. Rizzuto}}(2016)}]{ZWPR}%
  \BibitemOpen
  \bibfield  {author} {\bibinfo {author} {\bibnamefont {{W. Zhou, R. Passante
  and L. Rizzuto}}},\ }\bibfield  {title} {\enquote {\bibinfo {title}
  {{Resonance interaction energy between two accelerated identical atoms in a
  coaccelerated frame and the Unruh effect}},}\ }\href {\doibase
  10.1103/PhysRevD.94.105025} {\bibfield  {journal} {\bibinfo  {journal} {Phys.
  Rev. D}\ }\textbf {\bibinfo {volume} {94}},\ \bibinfo {pages} {105025}
  (\bibinfo {year} {2016})}\BibitemShut {NoStop}%
\bibitem [{\citenamefont {{F. Hong-Yi}}(2010)}]{FHY}%
  \BibitemOpen
  \bibfield  {author} {\bibinfo {author} {\bibnamefont {{F. Hong-Yi}}},\
  }\bibfield  {title} {\enquote {\bibinfo {title} {{Number-Conserving Coherent
  State in Rindler Space}},}\ }\href {\doibase 10.1088/0253-6102/54/3/15}
  {\bibfield  {journal} {\bibinfo  {journal} {Communications in Theoretical
  Physics}\ }\textbf {\bibinfo {volume} {54}},\ \bibinfo {pages} {457}
  (\bibinfo {year} {2010})}\BibitemShut {NoStop}%
\bibitem [{\citenamefont {{E. Mart\'in-Mart\'inez, I. Fuentes and R. B.
  Mann}}(2011)}]{MMFE}%
  \BibitemOpen
  \bibfield  {author} {\bibinfo {author} {\bibnamefont {{E.
  Mart\'in-Mart\'inez, I. Fuentes and R. B. Mann}}},\ }\bibfield  {title}
  {\enquote {\bibinfo {title} {{Using Berry's Phase to Detect the Unruh Effect
  at Lower Accelerations}},}\ }\href {\doibase 10.1103/PhysRevLett.107.131301}
  {\bibfield  {journal} {\bibinfo  {journal} {Phys. Rev. Lett.}\ }\textbf
  {\bibinfo {volume} {107}},\ \bibinfo {pages} {131301} (\bibinfo {year}
  {2011})}\BibitemShut {NoStop}%
\bibitem [{\citenamefont {{E. Mart\'in-Mart\'inez, A. Dragan, R. B. Mann and I.
  Fuentes}}(2013)}]{MMED}%
  \BibitemOpen
  \bibfield  {author} {\bibinfo {author} {\bibnamefont {{E.
  Mart\'in-Mart\'inez, A. Dragan, R. B. Mann and I. Fuentes}}},\ }\bibfield
  {title} {\enquote {\bibinfo {title} {{Berry phase quantum thermometer}},}\
  }\href {\doibase 10.1088/1367-2630/15/5/053036} {\bibfield  {journal}
  {\bibinfo  {journal} {New Journal of Physics}\ }\textbf {\bibinfo {volume}
  {15}},\ \bibinfo {pages} {053036} (\bibinfo {year} {2013})}\BibitemShut
  {NoStop}%
\bibitem [{\citenamefont {Fulling}(1973)}]{SAFNCQ}%
  \BibitemOpen
  \bibfield  {author} {\bibinfo {author} {\bibfnamefont {S.~A.}\ \bibnamefont
  {Fulling}},\ }\bibfield  {title} {\enquote {\bibinfo {title} {{Nonuniqueness
  of Canonical Field Quantization in Riemannian Space-Time}},}\ }\href
  {\doibase 10.1103/PhysRevD.7.2850} {\bibfield  {journal} {\bibinfo  {journal}
  {Phys. Rev. D}\ }\textbf {\bibinfo {volume} {7}},\ \bibinfo {pages}
  {2850--2862} (\bibinfo {year} {1973})}\BibitemShut {NoStop}%
\bibitem [{\citenamefont {Fulling}(1989)}]{FSA1}%
  \BibitemOpen
  \bibfield  {author} {\bibinfo {author} {\bibfnamefont {Stephen~A}\
  \bibnamefont {Fulling}},\ }\href@noop {} {\emph {\bibinfo {title} {Aspects of
  quantum field theory in curved space-time}}}\ (\bibinfo  {publisher}
  {Cambridge University Press},\ \bibinfo {year} {1989})\BibitemShut {NoStop}%
\bibitem [{\citenamefont {{F. Lenz, K. Ohta and K. Yazaki}}(2008)}]{LFOK}%
  \BibitemOpen
  \bibfield  {author} {\bibinfo {author} {\bibnamefont {{F. Lenz, K. Ohta and
  K. Yazaki}}},\ }\bibfield  {title} {\enquote {\bibinfo {title} {{Canonical
  quantization of gauge fields in static space-times with applications to
  Rindler spaces}},}\ }\href {\doibase 10.1103/PhysRevD.78.065026} {\bibfield
  {journal} {\bibinfo  {journal} {Phys. Rev. D}\ }\textbf {\bibinfo {volume}
  {78}},\ \bibinfo {pages} {065026} (\bibinfo {year} {2008})}\BibitemShut
  {NoStop}%
\bibitem [{\citenamefont {Svaiter}\ and\ \citenamefont {Svaiter}(1992)}]{SBS}%
  \BibitemOpen
  \bibfield  {author} {\bibinfo {author} {\bibfnamefont {B.~F.}\ \bibnamefont
  {Svaiter}}\ and\ \bibinfo {author} {\bibfnamefont {N.~F.}\ \bibnamefont
  {Svaiter}},\ }\bibfield  {title} {\enquote {\bibinfo {title} {Inertial and
  noninertial particle detectors and vacuum fluctuations},}\ }\href {\doibase
  10.1103/PhysRevD.46.5267} {\bibfield  {journal} {\bibinfo  {journal} {Phys.
  Rev. D}\ }\textbf {\bibinfo {volume} {46}},\ \bibinfo {pages} {5267--5277}
  (\bibinfo {year} {1992})}\BibitemShut {NoStop}%
\bibitem [{\citenamefont {Svaiter}\ and\ \citenamefont {Svaiter}(1993)}]{SBS1}%
  \BibitemOpen
  \bibfield  {author} {\bibinfo {author} {\bibfnamefont {B.~F.}\ \bibnamefont
  {Svaiter}}\ and\ \bibinfo {author} {\bibfnamefont {N.~F.}\ \bibnamefont
  {Svaiter}},\ }\bibfield  {title} {\enquote {\bibinfo {title} {Erratum:
  Inertial and noninertial particle detectors and vacuum fluctuations},}\
  }\href {\doibase 10.1103/PhysRevD.47.4802} {\bibfield  {journal} {\bibinfo
  {journal} {Phys. Rev. D}\ }\textbf {\bibinfo {volume} {47}},\ \bibinfo
  {pages} {4802--4802} (\bibinfo {year} {1993})}\BibitemShut {NoStop}%
\bibitem [{\citenamefont {Higuchi}\ \emph {et~al.}(1993)\citenamefont
  {Higuchi}, \citenamefont {Matsas},\ and\ \citenamefont {Peres}}]{HAM}%
  \BibitemOpen
  \bibfield  {author} {\bibinfo {author} {\bibfnamefont {A.}~\bibnamefont
  {Higuchi}}, \bibinfo {author} {\bibfnamefont {G.~E.~A.}\ \bibnamefont
  {Matsas}}, \ and\ \bibinfo {author} {\bibfnamefont {C.~B.}\ \bibnamefont
  {Peres}},\ }\bibfield  {title} {\enquote {\bibinfo {title} {Uniformly
  accelerated finite-time detectors},}\ }\href {\doibase
  10.1103/PhysRevD.48.3731} {\bibfield  {journal} {\bibinfo  {journal} {Phys.
  Rev. D}\ }\textbf {\bibinfo {volume} {48}},\ \bibinfo {pages} {3731--3734}
  (\bibinfo {year} {1993})}\BibitemShut {NoStop}%
\bibitem [{\citenamefont {Sriramkumar}\ and\ \citenamefont
  {Padmanabhan}(1996)}]{SLPT}%
  \BibitemOpen
  \bibfield  {author} {\bibinfo {author} {\bibfnamefont {L}~\bibnamefont
  {Sriramkumar}}\ and\ \bibinfo {author} {\bibfnamefont {T}~\bibnamefont
  {Padmanabhan}},\ }\bibfield  {title} {\enquote {\bibinfo {title}
  {{Finite-time response of inertial and uniformly accelerated Unruh - {DeWitt}
  detectors}},}\ }\href {\doibase 10.1088/0264-9381/13/8/005} {\bibfield
  {journal} {\bibinfo  {journal} {Classical and Quantum Gravity}\ }\textbf
  {\bibinfo {volume} {13}},\ \bibinfo {pages} {2061--2079} (\bibinfo {year}
  {1996})}\BibitemShut {NoStop}%
\bibitem [{\citenamefont {Lima}\ \emph {et~al.}(2019)\citenamefont {Lima},
  \citenamefont {Brito}, \citenamefont {Hoyos},\ and\ \citenamefont
  {Vanzella}}]{LCBF}%
  \BibitemOpen
  \bibfield  {author} {\bibinfo {author} {\bibfnamefont {Cesar A.~Uliana}\
  \bibnamefont {Lima}}, \bibinfo {author} {\bibfnamefont {Frederico}\
  \bibnamefont {Brito}}, \bibinfo {author} {\bibfnamefont {Jos{\'e}A.}\
  \bibnamefont {Hoyos}}, \ and\ \bibinfo {author} {\bibfnamefont {Daniel
  A.~Turolla}\ \bibnamefont {Vanzella}},\ }\bibfield  {title} {\enquote
  {\bibinfo {title} {{Probing the Unruh effect with an accelerated extended
  system}},}\ }\href {\doibase 10.1038/s41467-019-10962-y} {\bibfield
  {journal} {\bibinfo  {journal} {Nature Communications}\ }\textbf {\bibinfo
  {volume} {10}},\ \bibinfo {pages} {3030} (\bibinfo {year}
  {2019})}\BibitemShut {NoStop}%
\bibitem [{\citenamefont {Chen}\ and\ \citenamefont {Tajima}(1999)}]{CPT}%
  \BibitemOpen
  \bibfield  {author} {\bibinfo {author} {\bibfnamefont {Pisin}\ \bibnamefont
  {Chen}}\ and\ \bibinfo {author} {\bibfnamefont {Toshi}\ \bibnamefont
  {Tajima}},\ }\bibfield  {title} {\enquote {\bibinfo {title} {{Testing Unruh
  Radiation with Ultraintense Lasers}},}\ }\href {\doibase
  10.1103/PhysRevLett.83.256} {\bibfield  {journal} {\bibinfo  {journal} {Phys.
  Rev. Lett.}\ }\textbf {\bibinfo {volume} {83}},\ \bibinfo {pages} {256--259}
  (\bibinfo {year} {1999})}\BibitemShut {NoStop}%
\bibitem [{\citenamefont {{G. P. Berman, D. I. Kamenev and V. I.
  Tsifrinovich}}(2005)}]{BKT}%
  \BibitemOpen
  \bibfield  {author} {\bibinfo {author} {\bibnamefont {{G. P. Berman, D. I.
  Kamenev and V. I. Tsifrinovich}}},\ }\href@noop {} {\emph {\bibinfo {title}
  {{Perturbation Theory for Solid-state Quantum Computation With Many Quantum
  Bits}}}}\ (\bibinfo  {publisher} {Rinton Pr Inc.},\ \bibinfo {year}
  {2005})\BibitemShut {NoStop}%
\bibitem [{\citenamefont {{M. Hillery}}(1984)}]{HILLERY}%
  \BibitemOpen
  \bibfield  {author} {\bibinfo {author} {\bibnamefont {{M. Hillery}}},\
  }\bibfield  {title} {\enquote {\bibinfo {title} {{Distribution functions in
  physics: Fundamentals}},}\ }\href {\doibase 10.1016/0370-1573(84)90160-1}
  {\bibfield  {journal} {\bibinfo  {journal} {Phys. Rep.}\ }\textbf {\bibinfo
  {volume} {106}},\ \bibinfo {pages} {121 -- 167} (\bibinfo {year}
  {1984})}\BibitemShut {NoStop}%
\end{thebibliography}

%

\newpage

\appendix
 \begin{widetext}
\section{Supplemental material}

\subsection{Evolution operator}

To proceed with calculations of evolution operator, we use the interaction picture writing the 
interaction Hamiltonian as,
\begin{align}
H_{I}(\tau)= e^{-iH_f \tau}H_{sd} e^{iH_f \tau},
\end{align}
where $H_f  = \one \otimes\sum_{\omega > 0 
}\sum_{ \textbf{k}_{\bot}} 
\omega b^\dagger_{\mathbf k} b_{\mathbf k}  $ and,
\begin{align}
H_{sd} = {{\varepsilon}\over{2}}\sigma_z \otimes \one 
+\frac{\lambda}{\sqrt{V}}\sigma_z \otimes\sum_{\omega > 0 
}\sum_{ \textbf{k}_{\bot}}  \frac{1 }{\sqrt{2\omega}}(g_{\mathbf k} b^\dagger_{\mathbf k}+ 
g^\ast_{\mathbf k} b_{\mathbf k}).
\end{align}
Then, in the interaction picture, the evolution operator,  $U(\tau) =\hat T \exp \Big(- 
{i}\int_{0}^\tau dt' H_{I}(t')\Big )$,  
can be written as \cite{BKT},
\begin{align}
U(\tau)  = \exp 
\Big(-i\nu(\tau)  \one + i \frac{\varepsilon}{2}\sigma_z \otimes \one +\sigma_z 
\otimes\sum_{{\mathbf k}}  (b^\dagger_{\mathbf k} \xi_{\mathbf k}(\tau)- b_{\mathbf k} 
\xi^\ast_{\mathbf k}(\tau))\Big ),
\label{UO}
\end{align}	
where,
\begin{align}
\nu(\tau) =\frac{i}{2}\sum_{{\mathbf k}} \int^\tau_{0}\big(\xi^\ast_{\mathbf 
k}(t)\dot{\xi}_{\mathbf k}(t) - \xi_{\mathbf k}(t)\dot{\xi}^\ast_{\mathbf k}(t)\big ) dt, \quad
\xi_{\mathbf k}(\tau) =  \lambda g_{\mathbf k}\frac{1- e^{-i \omega \tau}}{ 
\sqrt{2V}\omega^{3/2} } .
\end{align}

The results of the action of  the evolution operator \eqref{UO} on an arbitrary pure initial state 
of the entire system, can  be described by the expressions,
\begin{align}
\label{D1a}
U(\tau) |\downarrow \rangle\otimes |\psi_F\rangle = 
&e^{-i\nu(\tau) -  i \varepsilon \tau/2}|\downarrow\rangle\otimes \prod_k D\Big 
(-\xi_{\mathbf 
k}(\tau)\Big 
)|\psi_F\rangle, \\
U(\tau) |\uparrow \rangle\otimes |\psi_F\rangle =& e^{-i\nu(\tau) + i \varepsilon 
\tau/2}|\uparrow 
\rangle\otimes \prod_k D\Big(\xi_{\mathbf k}(\tau) \Big)|\psi_F\rangle,
\label{D1b}
\end{align}
where $|\psi_F\rangle $ is an initial state of the scalar field, and $D(\xi_{\mathbf k} ) $ denotes  the displacement operator \cite{HILLERY}:
\begin{align}
D(\xi_{\mathbf k} ) =e^{\xi_{\mathbf k} b^\dagger_{\mathbf k}- \xi^\ast_{\mathbf k} b_{\mathbf k} }.
\label{EqD}
\end{align}

The density operator, $\varrho(\tau)$, for the entire system,  at time $\tau=\tau_0$,  is taken 
in the form, $\varrho(\tau_0) = |\Psi_0\rangle\langle\Psi_0 |$. Here,  
$|\Psi_0\rangle = |\psi_0\rangle\otimes |0_M \rangle$. The  wave function, $|\psi_0\rangle$,  
denotes the initial state of the detector, and $|0_M \rangle$ stands for the Minkowski vacuum. 
The reduced density matrix of the detector, $\rho(\tau)$, is obtained by tracing out all scalar 
field degrees of freedom. Its matrix elements are given by,
\begin{align} \label{EqR}
\rho_{ij}(\tau) =  \langle i| {\rm Tr}_R U(\tau)\varrho(\tau_0) U^{-1}(\tau)|j \rangle, 
\quad (i,j=0,1), 
\end{align}
where the index $i=0$ is associated with the eigenvector $|\downarrow \rangle$ and the index 
$i=1$ is associated with the eigenvector $|\uparrow \rangle$.
Using Eqs. (\ref{D1a}) and (\ref{D1b}), one can show that $\rho_{00}(\tau)=\rho_{00}(\tau_0)$, 
$\rho_{11}(\tau)=\rho_{11}(\tau_0),$ and,
\begin{align}
\rho_{01} (\tau)=  e^{  i \varepsilon \tau-\gamma(\tau)}\rho_{01} (\tau_0).
\end{align}

\subsection{ Bogolyubov transformations}

The relations between the creation and annihilation operators in the Rindler spacetime and 
in the Minkowski spacetime are determined by the Bogolyubov transformations:
$\hat b_{k} =\sum_p (\alpha_{{\mathbf k} {\mathbf p}}\hat a_{\mathbf p} - \beta^{\ast}_{{\mathbf k} {\mathbf p}}\hat a^\dagger_{\mathbf p})$.
Note that, while the  operators $\hat b_{\mathbf k}$  annihilate the Rindler vacuum, $|0_R \rangle$,  the operators $\hat a_{\mathbf p}$ annihilate the Minkowski vacuum, $|0_M \rangle$. The computation of the  Bogolyubov coefficients yields \cite{CLH}:
\begin{align}
\hat{b}_{\omega \mathrm{k}_{\perp}}=\frac{\hat{a}_{-\omega \mathbf{k}_{\perp}}+e^{-\pi \omega / a} \hat{a}_{\omega-\mathbf{k}_{\perp}}^{\dagger}}{\sqrt{1-e^{-2 \pi \omega / a}}}.
\label{BT}
\end{align}
Using the Bogolyubov transformations, one can easily calculate the mean number of the particles in the mode $\omega$: 
\begin{align}
\bar n_\omega =\langle 0_M |b^\dagger_{\mathbf k}b_{\mathbf k} |0_M\rangle  =\frac{1}{e^{2\pi \omega/a} -1}.
\end{align}

In order to obtain the decoherence function,  $\gamma(\tau)$, we employ the Bogolyubov 
transformations. Substituting \eqref{BT} in  Eq. \eqref{EqD} and using Eqs. \eqref{D1a}, 
\eqref{D1b} and \eqref{EqR}, after some algebra we obtain,
\begin{align}
\gamma(\tau)=  {2} \sum_{{\mathbf k}}|\xi_k(\tau)|^2 (1 +2\bar n_k) 
= {2} \sum_{{\mathbf k}}|\xi_k(\tau)|^2 \coth\Big(\frac{\pi \omega_k}{a}\Big).
\end{align}

In the continuous limit, the sum over ${\mathbf k} $ is replaced by the integral, $\sum 
\rightarrow V/(2\pi)^3 \int d^3 {\mathbf k}$, and the decoherence function is defined by the 
integral:
\begin{align}
\gamma (\tau)=\frac{\lambda^2}{4\pi^3} \int_0^\infty \frac{|g_\omega|^2}{\omega}{\rm 
sinc}^2 \Big(\frac{\omega \tau}{2} \Big )\coth\Big(\frac{\pi \omega}{a}\Big) d\omega,
\label{D5a}
\end{align}
where, $|g_\omega |^2= \int |g_{\mathbf k}|^2 d^2 {\mathbf k}_\perp $ and  ${\rm sinc} (x) 
=\sin (x)/x$. 

\end{widetext}
\end{document}